\documentclass[a4paper]{article}
\usepackage{graphicx}
\usepackage{INTERSPEECH2022}
\usepackage{cite}

\title{An objective test tool for pitch extractors' response attributes}
\name{Hideki Kawahara$^1$, Kohei Yatabe$^2$, Ken-Ichi Sakakibara$^3$, 
Tatsuya Kitamura$^4$, Hideki Banno$^5$,\\ Masanori Morise$^6$}
\address{
  $^1$Wakayama University, Wakayama, 640-8510 Japan\\
  $^2$Tokyo University of Agriculture and Technology, Tokyo, 184-8588 Japan\hspace{0mm}~\\
  $^3$Health Science University of Hokkaido, Hokkaido, 061-0293 Japan\hspace{0mm}~\\
  $^4$Konan University, Kobe 658-8510 Japan\hspace{0mm}~\\
  $^5$Meijo University, Nagoya 468-8502 Japan\hspace{0mm}~\\
  $^6$Meiji University, Tokyo, 164-8525 Japan\hspace{0mm}~}
\email{kawahara@wakayama-u.ac.jp, yatabe@go.tuat.ac.jp,
kis@hoku-iryo-u.ac.jp,t-kitamu@konan-u.ac.jp, banno@meijo-u.ac.jp, mmorise@meiji.ac.jp}

\begin{document}

\maketitle
\begin{abstract}
We propose an objective measurement method for pitch extractors' responses to frequency-modulated signals.
It enables us to evaluate different pitch extractors with unified criteria.
The method uses extended time-stretched pulses combined by binary orthogonal sequences.
It provides simultaneous measurement results consisting of the linear and the non-linear time-invariant responses and random and time-varying responses.
We tested representative pitch extractors using fundamental frequencies spanning 80~Hz to 800~Hz with 1/48 octave steps and produced more than 2000 modulation frequency response plots.
We found that making scientific visualization by animating these plots enables us to understand different pitch extractors' behavior at once.
Such efficient and effortless inspection is impossible by inspecting all individual plots.
%
The proposed measurement method with visualization leads to further improvement of the performance of one of the extractors mentioned above.
In other words, our procedure turns the specific pitch extractor into the best reliable measuring equipment that is crucial for scientific research.
We open-sourced MATLAB codes of the proposed objective measurement method and visualization procedure.
\end{abstract}

\section{Introduction}
Scientific investigations need reliable and accurate measuring equipment.
A pitch extractor serves as critical equipment for investigating voice fundamental frequency.
We introduce a method to objectively measure pitch extractors' performance regarding the frequency transfer function, total distortions, and signal-to-noise ratio.
We implemented the method using MATLAB and open-sourced.
We measured representative pitch extractors and reported raw data, scientific visualization movies and characterized pitch extractors on several performance maps.
These measurement and visualization led us to make one of the pitch extractors a promising tool for voice scientific research.

The contributions of this article are as follows.
We introduced a new objective measurement method of pitch extractors which provides useful supplemental information to existing evaluation measures.
We applied this method on existing pitch extractors and tuned NINJAL extractor\cite{HidekiKawahara2017,kawahara2017accurate} a reliable and accurate measuring equipment for speech science research. %

\section{Backgound}
\label{sec:background}
Revisiting a research topic a quarter-century ago motivated this research.
The first author introduced an altered auditory feedback technique\footnote{We use the term ``altered auditory feedback'' here because it is common practice now. We used the term ``transformed auditory feedback'' to represent our paradigm.} and reported that our voice fundamental frequency ($f_\mathrm{o}$)\footnote{We use $f_\mathrm{o}$ (pronunciation ``ef oh'') representing the fundamental frequency\cite{titze2015jasaforum} instead of using conventional symbols such as F0.} responds to feedback pitch modification\cite{kawahara1994interactions}.
Despite decades of research, altered feedback still is a hot topic for investigating speech chain and underlying neural basis\cite{BEHROOZMAND2015418,PATEL2016772.e33,larson2016sensory,behroozmand2020modulation,peng2021causal}.
Those research focused on voluntary responses represented by the pitch shift paradigm and adaptation paradigm\cite{larson2005Jasa,houde2013PNAS}.\footnote{For detailed historical background and discussions on altered feedback research and relation to CAPRICEP-based method, refer\cite{kawahara2021interspeech,kawahara2021APSIPA}.}

Introduction of CAPRICEP-based method\cite{kawahara2021icassp} and voice pitch response to auditory stimulation, which is not an altered feedback voice, opened a new possibility for investigating less studied involuntary response\cite{kawahara2021interspeech,kawahara2021APSIPA}.
The experiment requires the measurement of $f_\mathrm{o}$ of periodic sounds without introducing nonlinearities and glitches for measuring voice response to auditory stimulation.
This requirement led us to propose an objective response measurement method\footnote{The proposed experimental procedure uses a new system response analysis method (CAPRICEP-based method afterward. It stands for Cascaded All-Pass filters with RandomIzed CEnter frequencies and Phase polarity\cite{kawahara2021icassp}). 
The CAPRICEP-based method simultaneously measures the linear time-invariant (LTI) response, the non-linear time-invariant (non-LTI) response, and random and time-varying responses.} of pitch extractors' response to frequency-modulated tones\footnote{Strictly, using ``pitch'' to represent $f_\mathrm{o}$\cite{plack2006pitch} is misleading. What we perceive is ``pitch,'' a psychological attribute, which highly correlates with $f_\mathrm{o}$. However, we do not distinguish the use of pitch and $f_\mathrm{o}$ in this article unless it introduces confusion.}. 
We assigned the modulation signal as input and the extracted pitch of the produced voice as output and fed them to the proposed response analysis method.
%
%
We found that this assignment enables us to acquire pitch extractors' behavior objectively and in detail.

\section{Principles of operation}
\label{sec:method}
Pitch extraction has long research history\cite{hess2012pitch,noll1967jasa,atal1972automatic,kawahara1999spcom,de2002yin,kawahara05_interspeech,drugman11interspeech} and still is a hot topic in speech processing\cite{degottex2014covarep,morise17b_interspeech,kim2018crepe}.
We propose to apply the CAPRICEP-based method\cite{kawahara2021icassp} that is adopted to investigate voice pitch response to auditory stimulation\cite{kawahara2021interspeech,kawahara2021APSIPA} to measure the responses of pitch extractors. 

\begin{figure}[tbp]
\begin{center}
\includegraphics[width=0.87\hsize]{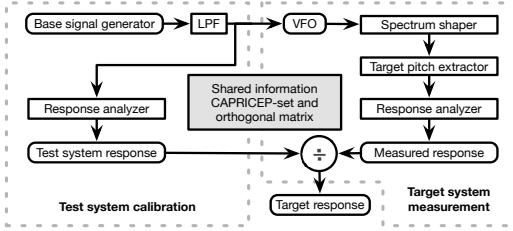}\\
\vspace{-3.5mm}
\caption{Measurement Scheme of the proposed method. (VFO: Variable Frequency Oscillator with harmonic components. The fundamental carrier frequency is the average $f_\mathrm{o}$.)}
\label{fig:measurementScheme}
\end{center}
\vspace{-8.0mm}
\end{figure}
\begin{figure}[tbp]
\begin{center}
\includegraphics[width=0.95\hsize]{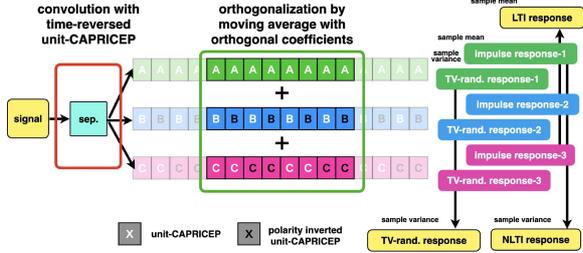}\\
\vspace{-2.5mm}
\caption{Schematic diagram of the response analyzer of Fig.~\ref{fig:measurementScheme}.}
\label{fig:capricepProcess}
\end{center}
\vspace{-9mm}
\end{figure}
Figure~\ref{fig:measurementScheme} shows the scheme of the proposed method.
The combination of elements \textbf{\textsf{Base signal generator}}, \textbf{\textsf{Response analyser}}, and \textbf{\textsf{CAPRICEP-set and orthogonal matrix}} is the CAPRICEP-based method\cite{kawahara2021icassp}.
We used a Gaussian smoother\footnote{We used a sidelobe-less Gaussian window by setting the endpoint level to the machine epsilon.} for \textbf{\textsf{LPF}} and fed the output to \textbf{\textsf{VFO}} (Variable Frequency Oscillator) to modulate $f_\mathrm{o}$ of the test signals.
We normalize the pitch extractor's responses (\textbf{\textsf{Target system measurement}}) by the test system response (\textbf{\textsf{Test system calibration}}) to get the \textbf{\textsf{Target response}}.
The pitch extractor's response should only consist of the LTI response and not introduce the non-LTI or random and time-varying responses\footnote{Strictly, this procedure introduces the non-LTI response. It is because FM is a non-linear operation, and its sideband is not bandlimited.}.

\subsection{Implementation}
\label{sec:imple}
Figure~\ref{fig:capricepProcess} shows a schemetic diagram of the response analyzer of Fig.~\ref{fig:measurementScheme}.
The test signal is a mixture of three sequences.
Convolution with three unit-CAPRICEP signals with time reversal (``A'', ``B'', and ``C'' in Fig.~\ref{fig:capricepProcess}) separates the test signal into three sequences.
The following orthogonalization procedure removes cross-correlation between sequences.
The LTI-response is the average of all element responses because the impulse response is identical irrespective of the test signals.
The difference between impulse responses obtained in each sequence provides sample variance representing random responses.
The difference between impulse responses obtained in different test sequences provides sample variance representing non-linear time-invariant response.
Please refer to details for~\cite{kawahara2021APSIPA}\footnote{We derived four times longer impulse response than each repetition length by using 1/4, 1/4, and 1/2 weight for the first, second, and third sequences and added. We used simplified description in the body text and, Fig.~\ref{fig:capricepProcess} for the actual calculation is too intricate.}.

We refined the implementation of algorithms described in~\cite{kawahara2021APSIPA} by taking advantage of the signal periodicity using FFT to calculate discrete Fourier transform.
Figure~\ref{fig:capricepProcess} shows that the test signal is a periodic signal with the length of the period $4N_u$, where $N_u$ is the allocation interval of each unit-CAPRICEP.
We made periodic signals $\tilde{x}_{k}[n]$ and $\tilde{y}_{k}[n]$ (where, $k \in \{1, 2, 3\}$, and $n = 0, \ldots, 4N_u -1$ represents the discrete time) from the separated test signals $x_{k}[n + n_0]$ and the measured signals $y_{k}[n + n_0]$ using the following equation.
\begin{align}
\tilde{x}_{k}[n] = & w[n] x_{k}[n + n_0] \nonumber \\
 & {} + w[4N_u - 1 - n]x_{k}[n + 4N_u + n_0] \label{eq:refPer} \\
\tilde{y}_{k}[n] = & w[n] y_{k}[n + n_0] \nonumber \\
 & {} + w[4N_u - 1 - n]y_{k}[n + 4N_u + n_0] , \label{eq:mesPer}
\end{align}
where $w[n]$ represents a weighting function that satisfies $w[n]+w[4N_u-1-n] = 1$, and $n_0$ represents the initial position of the selected signal segment\footnote{We used the integrated and normalized rectangular windowing function to remove level discontinuity of the truncated observation noise. This weighting needs 1.76~dB compensation of the calculated sample variance. Integrated and normalized half-wave rectified cosine windowing function assures continuity up to the first-order derivative with 1.25~dB compensation.}.

\subsection{Interface function for pitch extractors}
\label{sec:wrapper}
We prepared an interface function for each pitch extractor and called it from the response analyzer.
This interface makes it easy to test different pitch extractors.

\section{Measurement of pitch extractors}
\label{sec:mes}
We used MATLAB to implement the response analyzer and interface functions.
We tested representative pitch extractors to illustrate the proposed objective measurement method.

\subsection{Tested extractors}
\label{sec:testedEx}
The tested pitch extractors are listed below with abbreviations for representing them in plots.
Some extractors' test results are in supplement materials due to the limited available space.

\begin{description}
\item[MATLAB function (LHS,CEP,SRH,NCF,PEF,CREPE) ]
Scientific computing environment MATLAB has builtin functions for pitch extraction.
They consists of classical methods (``CEP'' cepstrum-based method\cite{noll1967jasa}, ``NCF'' LPC-based method\cite{atal1972automatic}, and ``LHS'' harmonic summation-based method\cite{hermes1988measurement}), and recent methods (``PEF'' \cite{Gonzalez2014PEFAC} and ``SRH'' summation of residual harmonics\cite{drugman11interspeech}).
In addition to these ``Deep Leaning Toolbox'' has a deep learning-based method ``CREPE''\cite{kim2018crepe}.
\item[YIN ]
YIN is an absolute difference-based method\cite{de2002yin}\footnote{%
The original YIN is compatible with the latest MATLAB version (R2022a) on Windows11. However, only one specific calling sequence of YIN (\texttt{r = yin(x,p)}) worked adequately on macOS.}.
\item[SWIPEP ]
SWIPEP uses a harmonic model-based approach.
It estimates the fundamental frequency of the best matching sawtooth signal\cite{camacho2008jasa}.
\item[RAPT (VOICEBOX) ]
RAPT uses a two-stage autocorrelation-based method followed by post processing\cite{talkin1995robust} for robust processing.
We used implementation in voicebox\cite{brucal2018female}.
\item[REAPER ] 
REAPER simultaneously estimate GCI (Glottal Closure Instant), V/UV (voiced or unvoiced), and pitch.
We used the open-source implementation\cite{googleREAPERgit}.
\item[Praat ]
Praat is a popular tool for doing phonetics using computers\cite{boersma2011praat}.
We used recommended procedure ``Sound: To Pitch...'' which uses the autocorrelation-based method\cite{boersma1993accurate} with the default setting.
\item[openSMILE ]
openSMILE is widely applied in automatic emotion recognition for affective computing\cite{eyben2010opensmile}. 
It has two configuration files for pitch analysis, ``prosodyShs.conf'' which uses the Sub-Harmonic-Sampling (SHS) method, and ``prosodyAcf.conf'' which uses an autocorrelation and cepstrum-based method (ACF).
\item[STRAIGHT (NDF, XSX) ]
Two VOCODER packages, legacy-STRAIGHT\cite{kawahara1999spcom} and TANDEM-STRAIGHT\cite{kawahara2008icassp} use $f_\mathrm{o}$ adaptive spectral envelope recovery.
Their $f_\mathrm{o}$ adaptive procedure led to development and refinement of dedicated pitch extractors, NDF\cite{kawahara05_interspeech} for legacy-STRAIGHT, and XSX\cite{kawahara08_spkd} for TANDEM-STRAIGHT.
\item[WORLD (Harvest) ]
A high-quality VOCODER WORLD\cite{morise2016world} also use $f_\mathrm{o}$ adaptive spectral envelope recovery.
The latest pitch extractor for WORLD is Harvest\cite{morise17b_interspeech}. 
\item[NINJAL ]
This pitch extractor\cite{HidekiKawahara2017,kawahara2017accurate} is a refined version of its predecessor\cite{Kawahara2016iscaSSW9}.
They use a log-linear filter-bank and their instantaneous frequency and residual levels of outputs.
We set the smoothing length parameter to 10~ms (named NINJALX2) and 40~ms (named NINJAL: default).
The setting of NINJALX2 is the result of fine-tuning enabled by the proposed objective measurement.
\end{description}

\subsection{Test signal generation and pitch extraction}
\label{sec:testcond}
We generated test signals using a 44100~Hz sampling rate.
For pitch extractors implemented using MATLAB, we fed the test signal directly.
We used a WAVE format file with 24~bit quantization for feeding the test signal to other pitch extractors.
The length of the signals is 20~s.
We set the fundamental frequencies of the test signals from 80~Hz to 800~Hz with 1/48 octave step.
We generated 1000 unit-CAPRICEP signals and selected a set consisting of 10 signals having minimum cross-correlation.

\begin{figure}[tbp]
\begin{center}
\includegraphics[width=0.85\hsize]{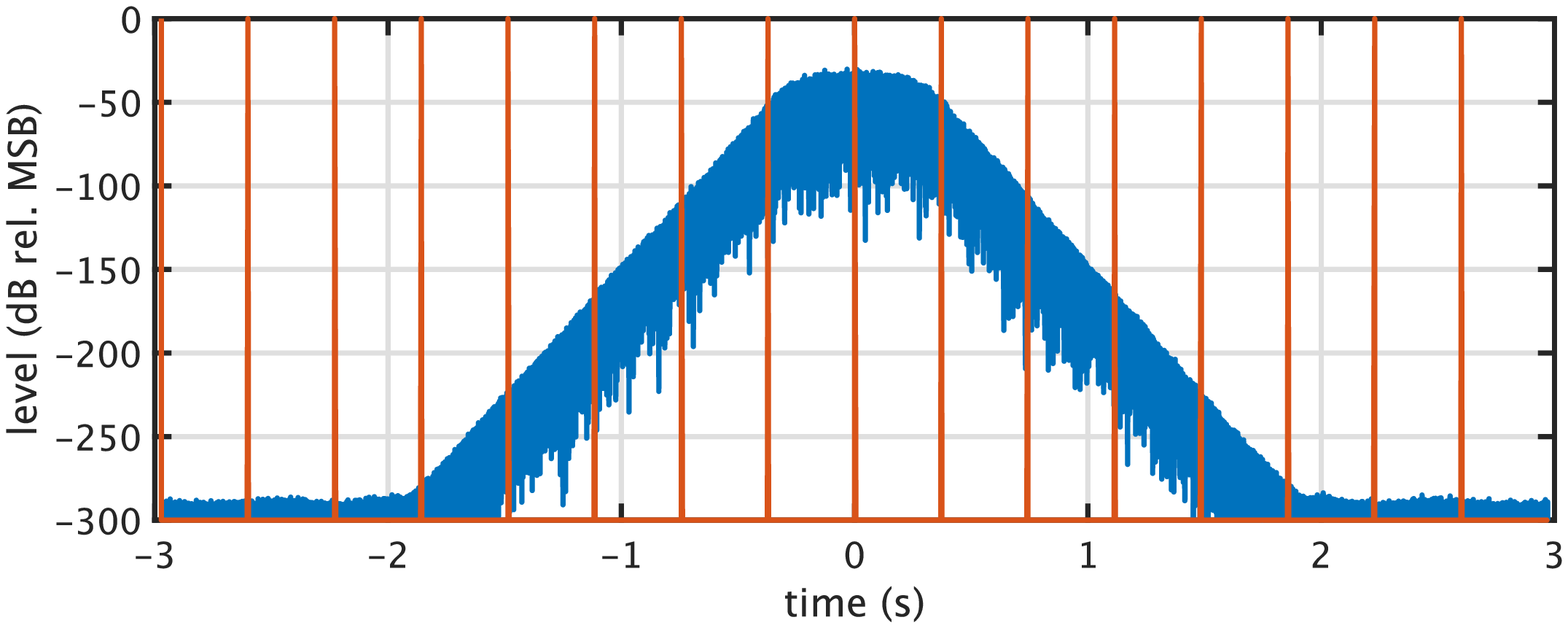}\\
\includegraphics[width=0.85\hsize]{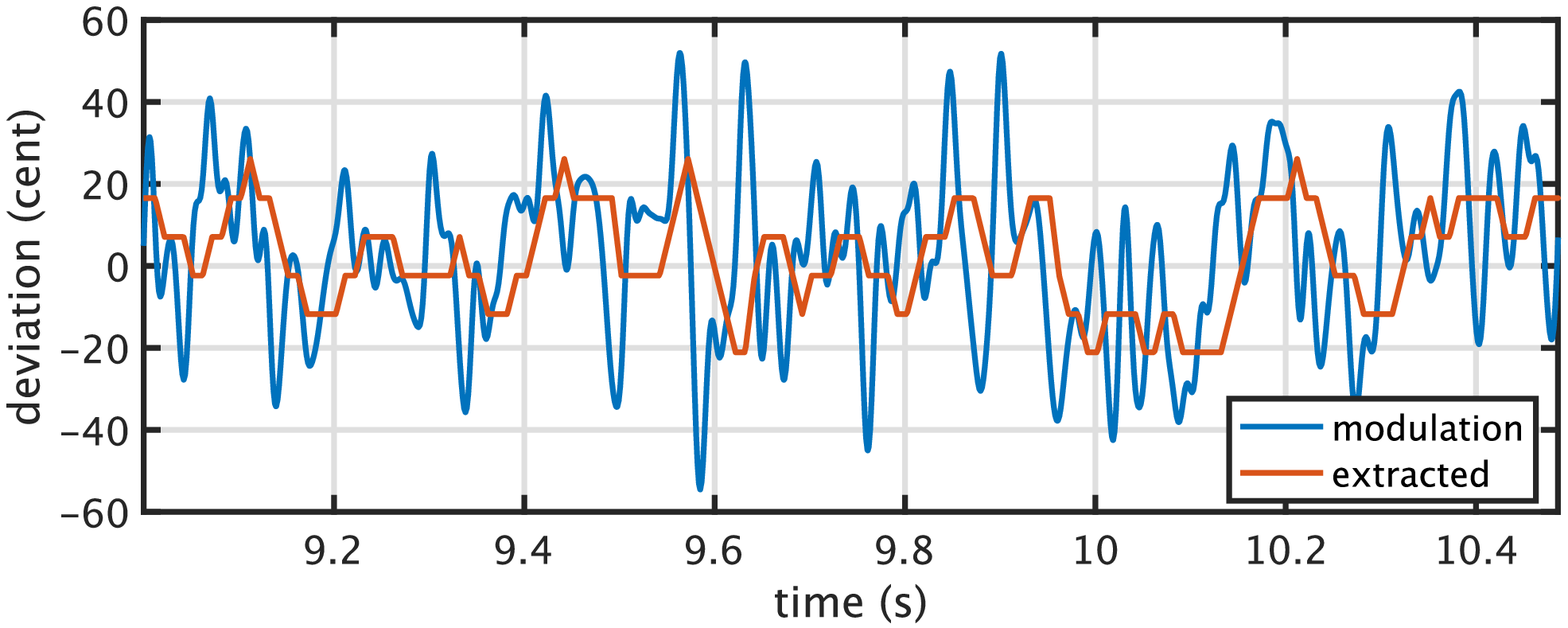}\\
\vspace{-2mm}
\caption{Upper plot: An example absolute value of unit-CAPRICEP.
The vertical lines represent the allocation interval.
Lower plot: frequency modulation signal (blue line) and the extracted pitch by NCF (red line).}
\label{fig:capricepAllocation}
\end{center}
\vspace{-6mm}
\end{figure}

The upper plot of Fig.~\ref{fig:capricepAllocation} shows an example of the absolute amplitude and allocation interval.
We selected three elements from the set of unit-CAPRICEP and allocated them 36 times with an overlap-and-add procedure changing polarity as shown in Fig.~\ref{fig:capricepProcess}.

We convolved the signal with a Gaussian shape and used it for modulating $f_\mathrm{o}$ of multiple harmonic signals\footnote{We applied modulation in the logarithmic frequency domain (musial cent) regarding $f_\mathrm{o}$ dynamics\cite{fujisaki1984analysis}.}.
Then we applied a spectral shaper simulating a Japanese vowel /a/.
The lower plot of Fig.~\ref{fig:capricepAllocation} shows an example of the generated modulation signal and example output of a pitch extractor NCF.
Note that the estimated pitch contour shows a quantization effect due to the sampling interval of discrete-time signals.

\subsection{Analysis example}
\label{sec:analysisEx}
We selected consecutive segments consisting of practically periodic signals.
First, we recovered the pulse by convolving time-reversed versions of each unit-CAPRICEP.
Then, we calculated the standard deviation of the difference of consecutive segments.
The length of each segment is $4N_u$, the fundamental period of the modulation signal.

\begin{figure}[tbp]
\begin{center}
\includegraphics[width=0.85\hsize]{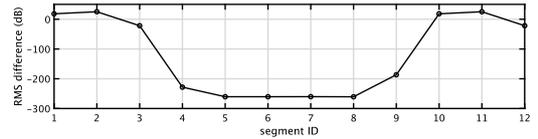}
\vspace{-3mm}
\caption{Standard deviation of the difference between consecutive segments.}
\label{fig:segmentSelection}
\end{center}
\vspace{-6mm}
\end{figure}
Figure~\ref{fig:segmentSelection} shows the standard deviations of differences between consecutive segments.
We set a -150~dB threshold level for selecting practically periodic pairs and selected six pairs.
Then, we made six periodic segments from the reference and the measured segments using Eq.~\ref{eq:refPer} and Eq.~\ref{eq:mesPer}.

\begin{figure}[tbp]
\begin{center}
\includegraphics[width=0.85\hsize]{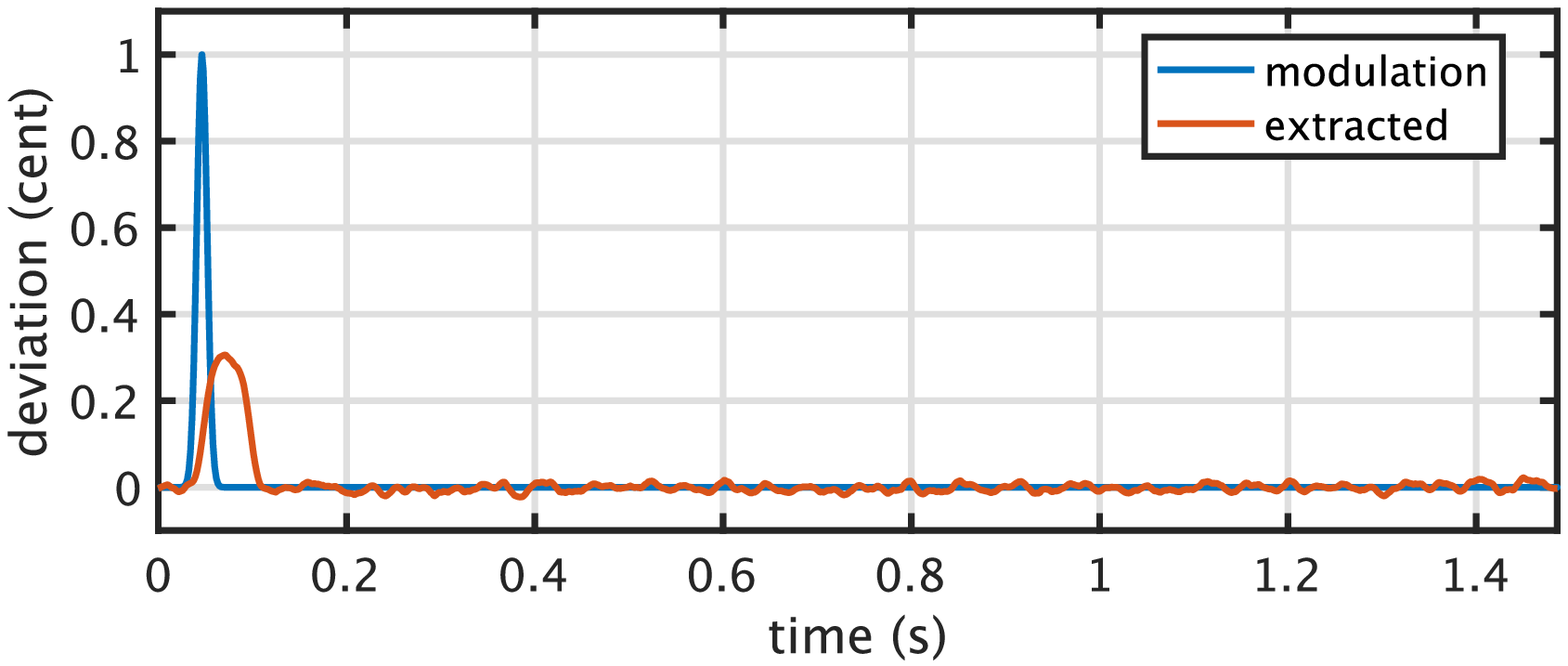}\\
\includegraphics[width=0.85\hsize]{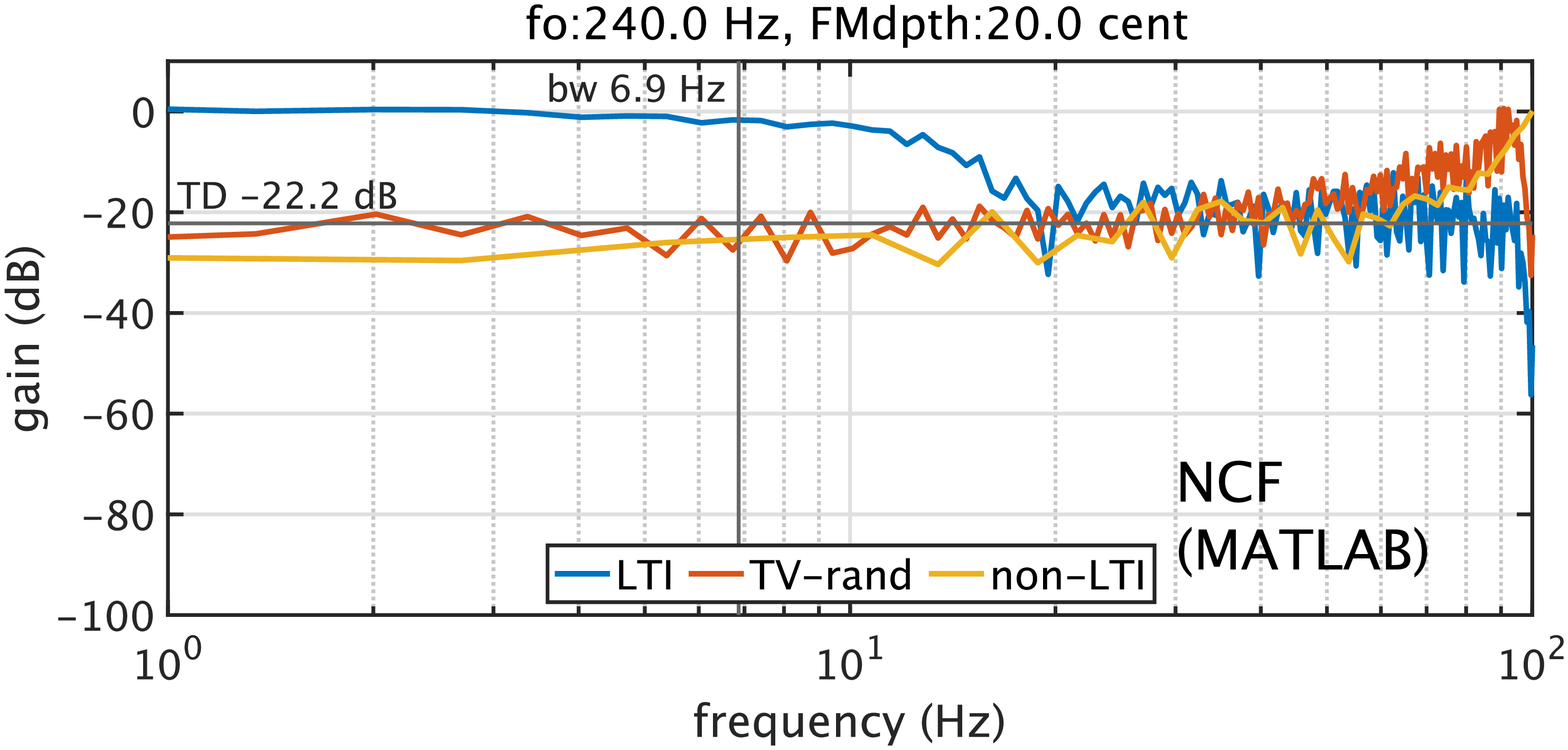}\\
\vspace{-3mm}
\caption{Upper plot: Applied modulation (blue line) and the impulse response (red line) of the NCF pitch extractor.
Bottom plot: Frequency characteristics of the LTI response (blue line), non-linear TI response (yellow line), and time-varying random response (red line). The vertical line represents the bandwidth (bw), and the horizontal line represents the total distortion (TD: average power of the random and non-LTI responses below bw.)}
\label{fig:testsignalprocess}
\end{center}
\vspace{-8mm}
\end{figure}
We applied the procedure in Fig.~\ref{fig:capricepProcess} and got 72 ($3 \times 4 \times 6$) impulse responses (length: $N_u$).
We used them to get six extended impulse responses (length: $4N_u$), average extended impulse response, and random and nonlinear responses.
The upper plot of Fig.~\ref{fig:testsignalprocess} shows examples of individual extended impulse response and the corresponding impulsive modulation.
The lower plot of Fig.~\ref{fig:testsignalprocess} shows the frequency characteristics of the responses.

\subsection{Response of representative extractors}
\label{sec:testres}
This section shows the frequency response of representative pitch extractors.
We placed the other extractors' responses in supplemental materials.

\begin{figure}[tbp]
\begin{center}
\includegraphics[width=0.74\hsize]{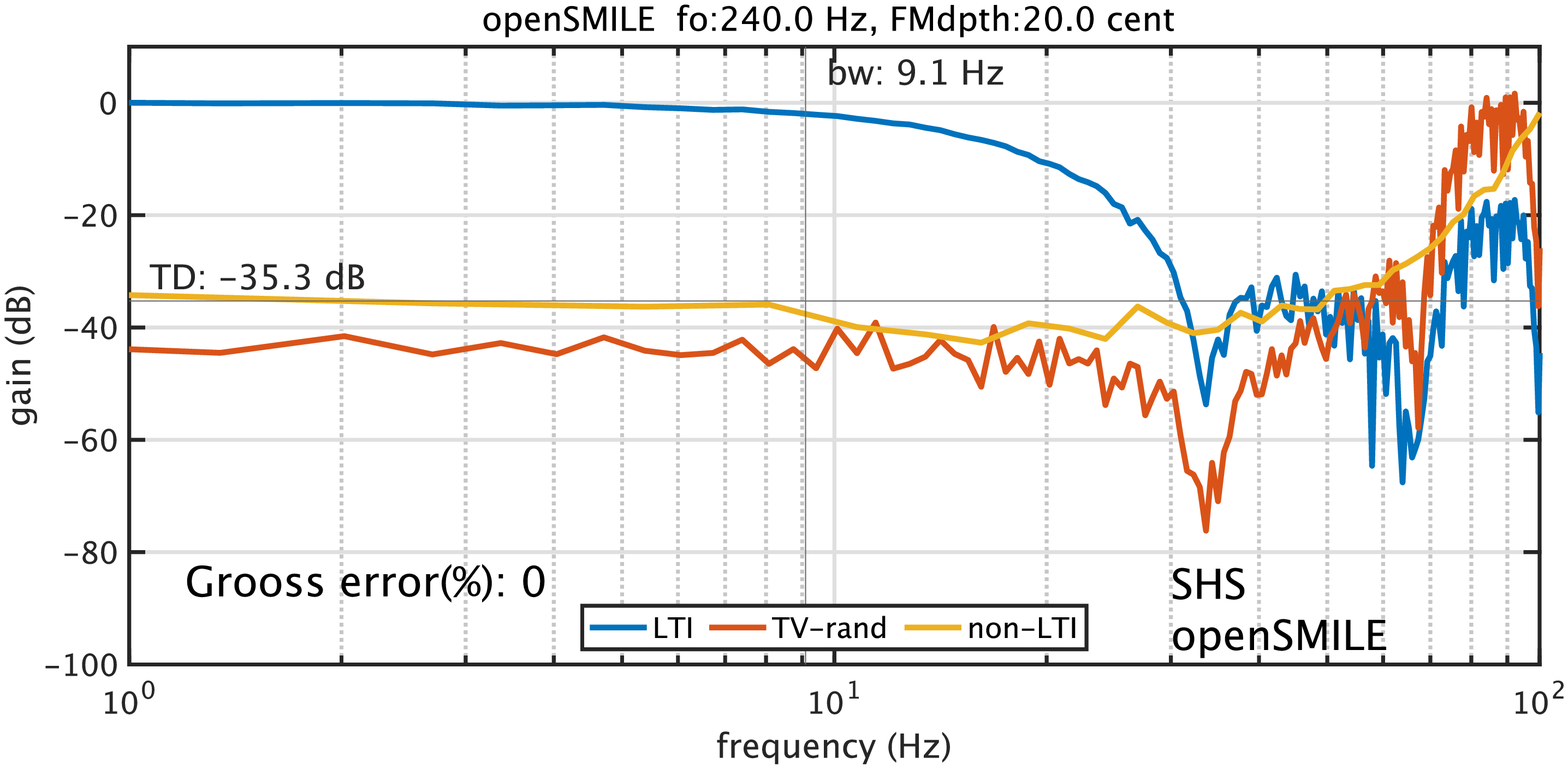}\\
\vspace{-3mm}
\includegraphics[width=0.74\hsize]{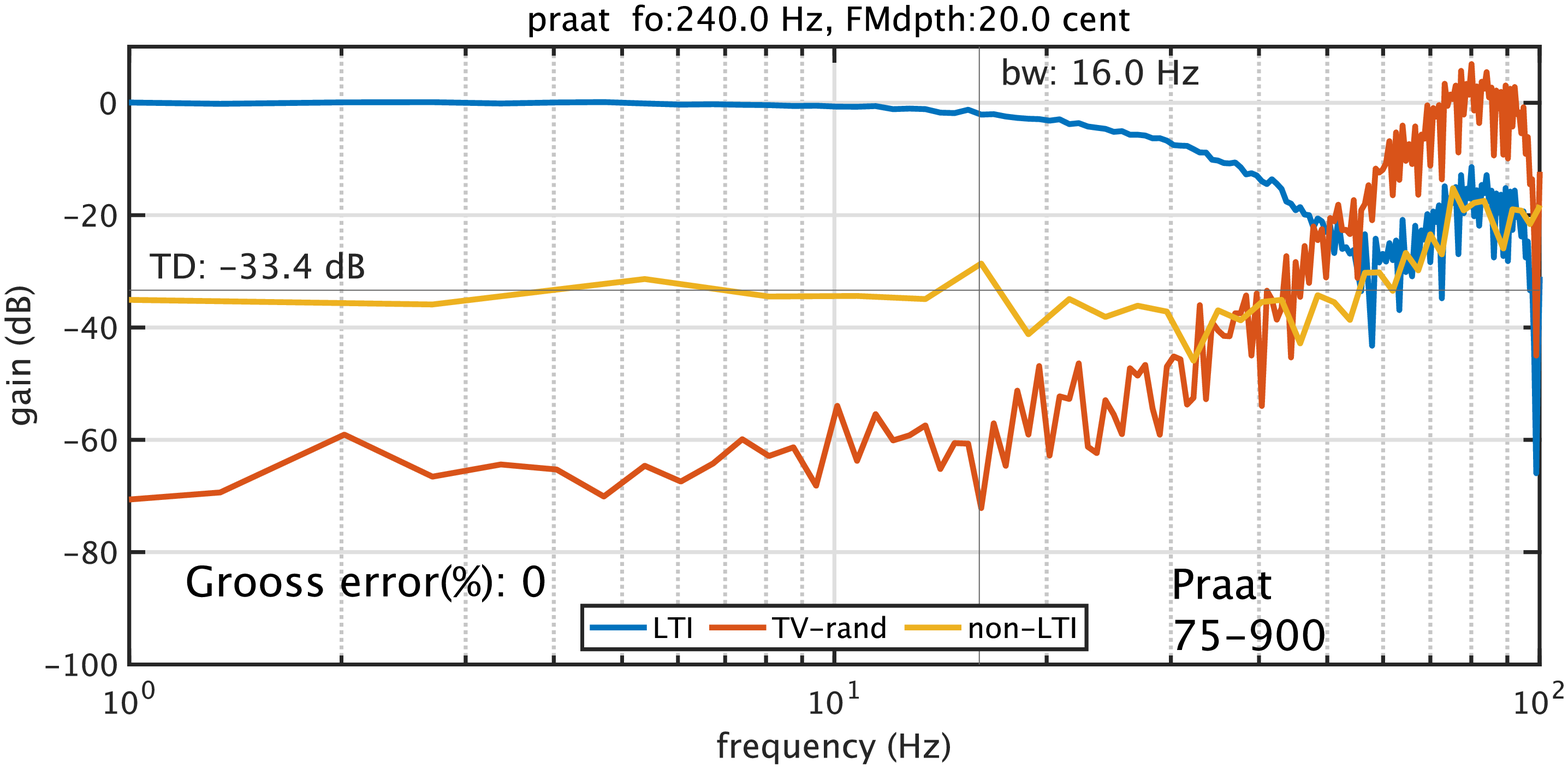}\\
\vspace{-3mm}
\includegraphics[width=0.74\hsize]{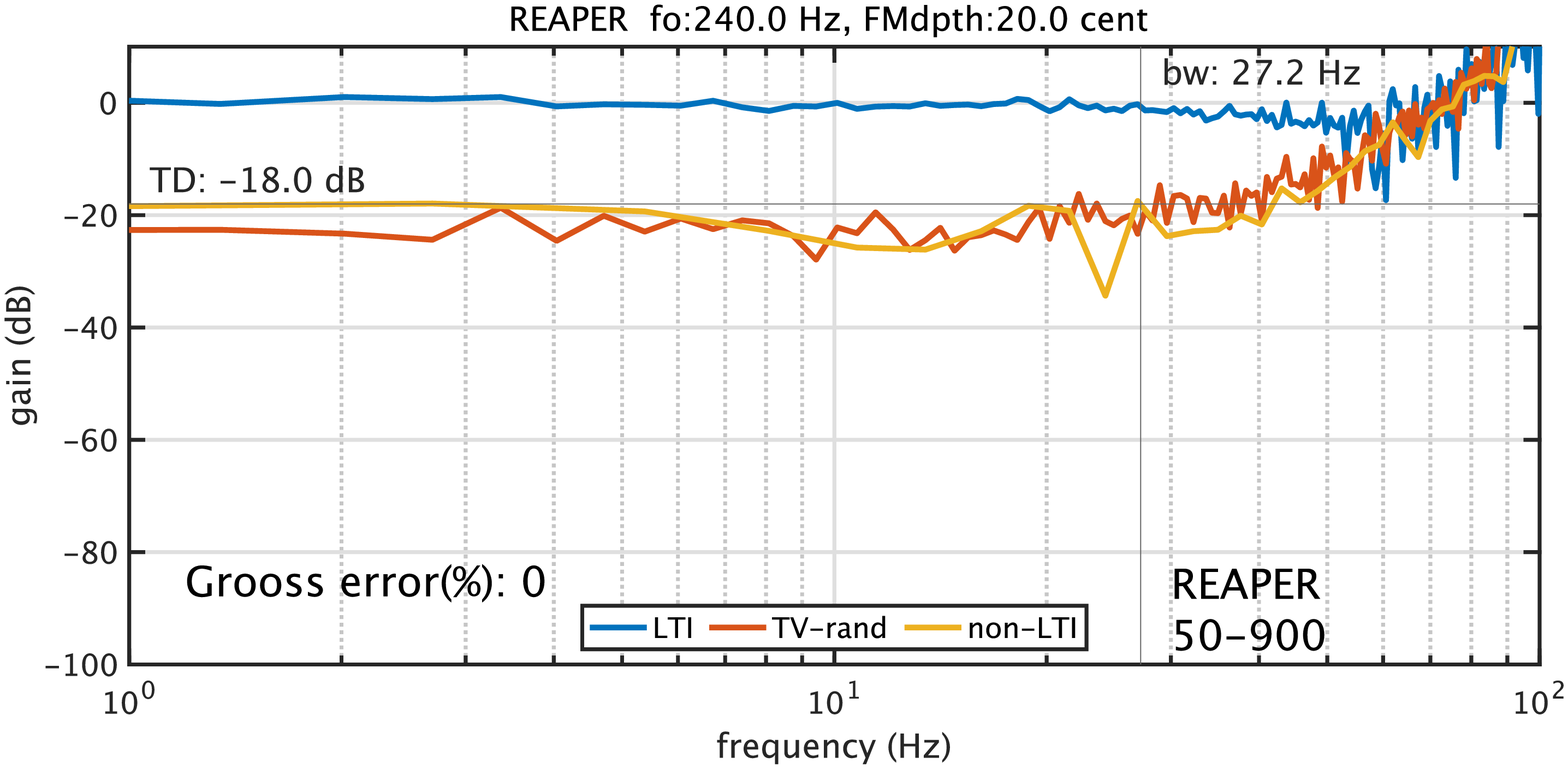}\\
\vspace{-3mm}
\caption{Frequency responses of pitch extractors (1).}
\label{fig:praat240Hz20220325T020838}
\end{center}
\vspace{-7mm}
\end{figure}
Figure~\ref{fig:praat240Hz20220325T020838} shows responses of popular pitch extractors.
SHS configuration of openSMILE provided better performance than ACF configuration.
Its default frame interval (10~ms) is the main cause of the narrow bandwidth.
Praat has wider bandwidth and relatively good total distortion.
REAPER has a significantly wide bandwidth.
However, the total distortion is poor.
This distortion is mainly because of the frequency quantization due to the sampling interval of the signal.

\begin{figure}[tbp]
\begin{center}
\includegraphics[width=0.74\hsize]{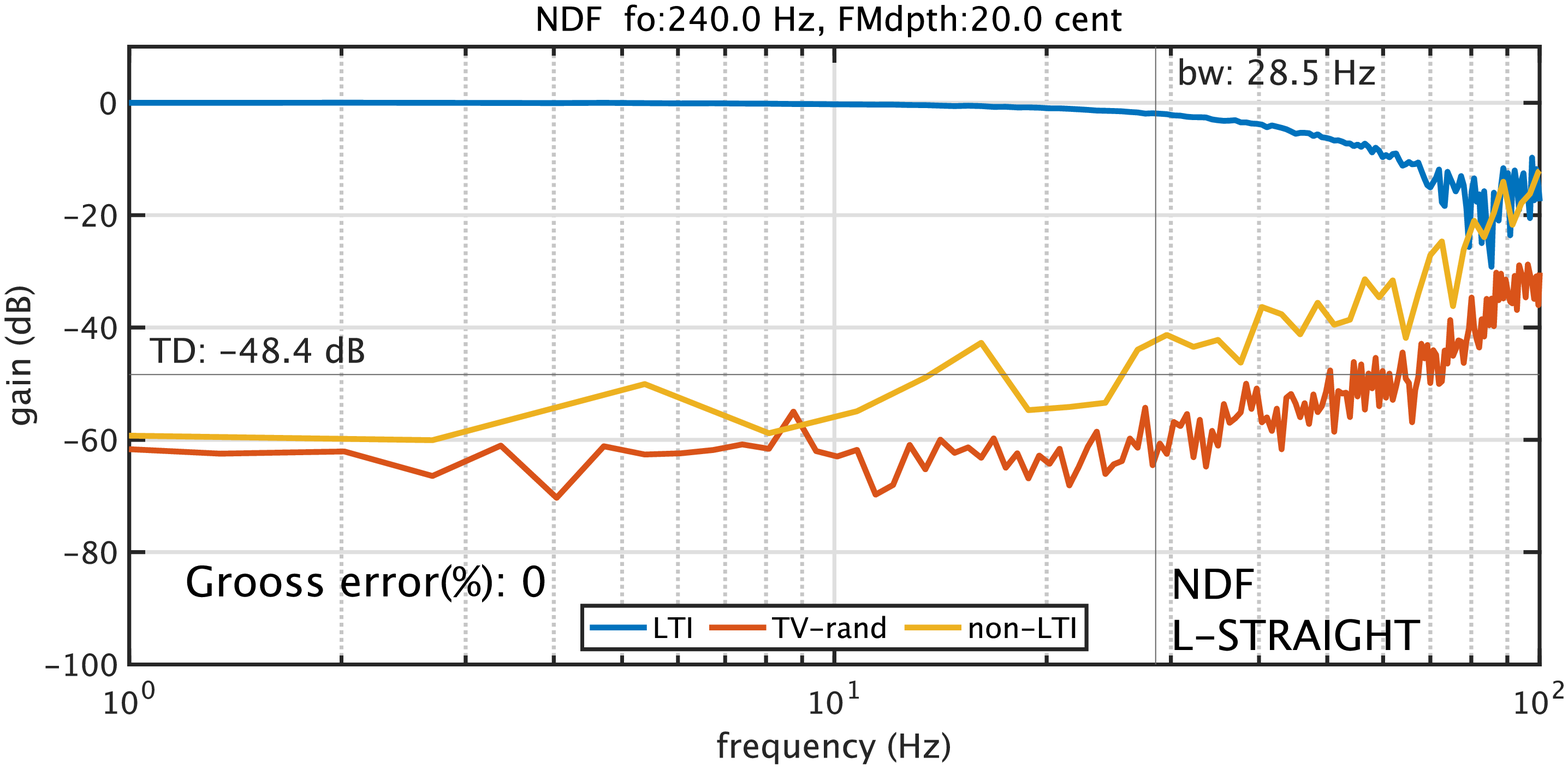}\\
\vspace{-3mm}
\includegraphics[width=0.74\hsize]{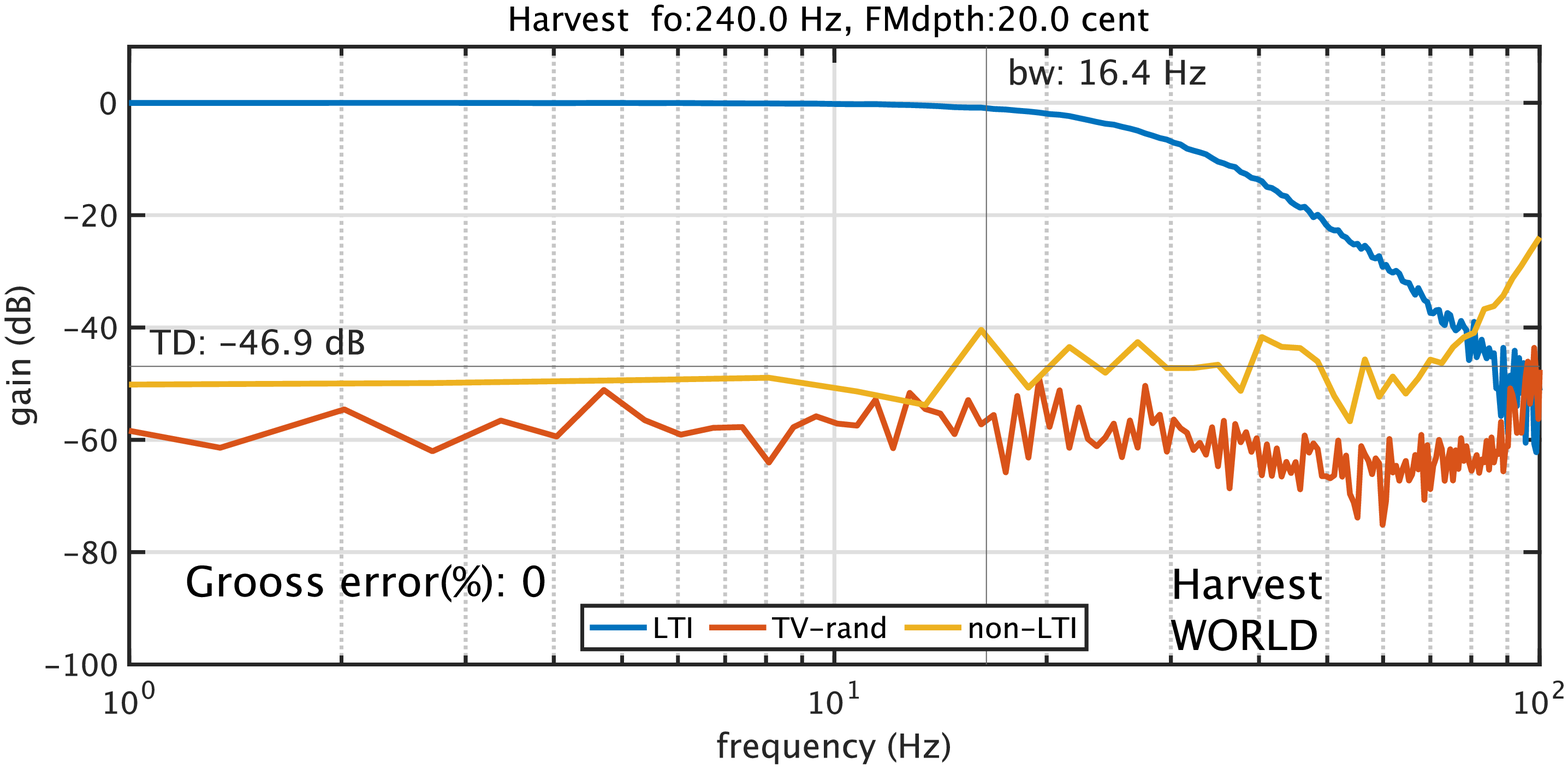}\\
\vspace{-3mm}
\includegraphics[width=0.74\hsize]{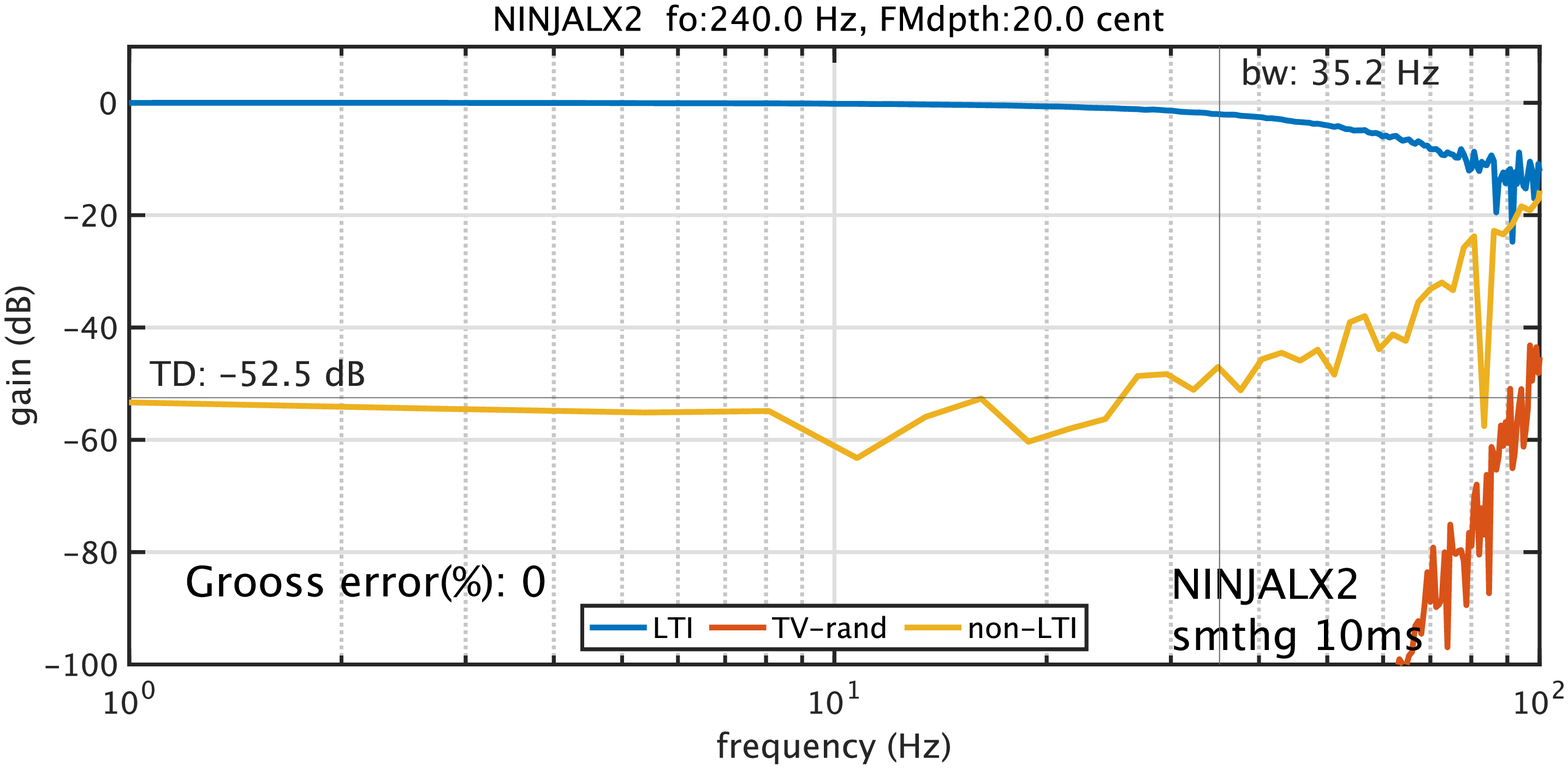}\\
\vspace{-3mm}
\caption{Frequency responses of pitch extractors (2).}
\label{fig:pNINJALX2240Hz20220325T021319}
\end{center}
\vspace{-9mm}
\end{figure}
Pitch extractors in Fig.~\ref{fig:pNINJALX2240Hz20220325T021319} show better performance both in the bandwidth and the total distortion than other extractors tested.
The significantly lower random response in NINJALX2 is due to its output rate (44100/6~Hz this case\footnote{It can operate at 44100~Hz. The default setting introduces automatic downsampling for reducing the processing time.}).
The relatively narrow bandwidth of Harvest is the result of tuning to avoid perceptually annoying pitch errors.

\subsection{Visualization and performance map}
\label{sec:visAnime}
\begin{figure}[tbp]
\begin{center}
\includegraphics[width=0.99\hsize]{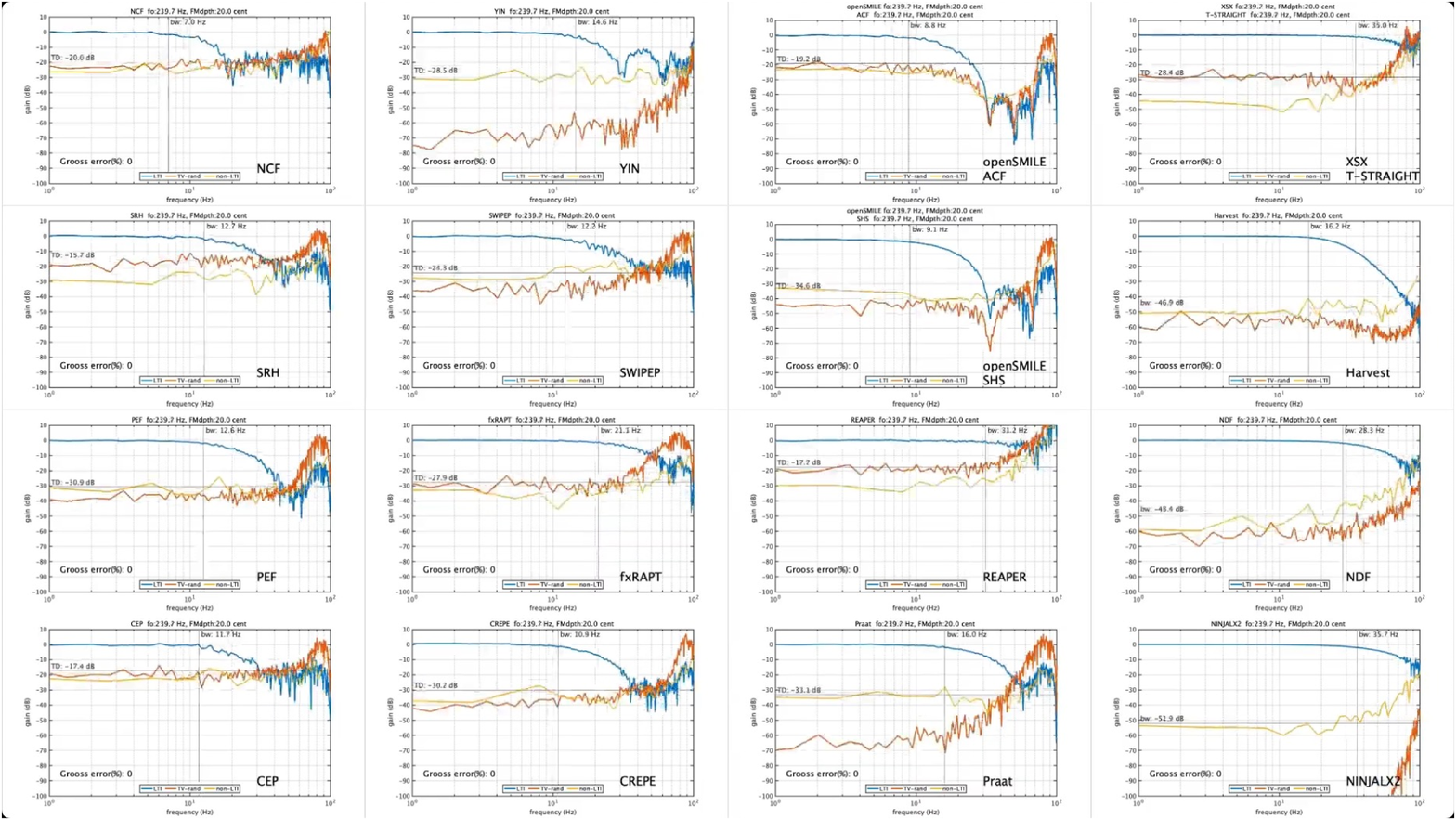}\\
\vspace{-2mm}
\caption{Snapshot of a movie frame scanning pitch from 80~Hz to 800~Hz in 5~seconds with a 1/48 octave step.}
\label{fig:movieSnapShot120}
\end{center}
\vspace{-6mm}
\end{figure}

The tests produced more than 2000 plots.
It was challenging to understand their behavior by inspecting all plots.
Instead, we compiled a movie of about 5 seconds, placing 16 plots having the same fundamental frequency in each movie frame.
Figure~\ref{fig:movieSnapShot120} shows a snapshot.
We found that movie visualization makes it easy to characterize the behavior of each extractor as a whole.
Please check the media file in the supplement. 

These visualizations suggested an improvement of our recent extractor, NINJAL.
We found the smoothing time of the original NINJAL is excessive by inspecting the movie and plots.
\begin{figure}[tbp]
\begin{center}
\includegraphics[width=0.99\hsize]{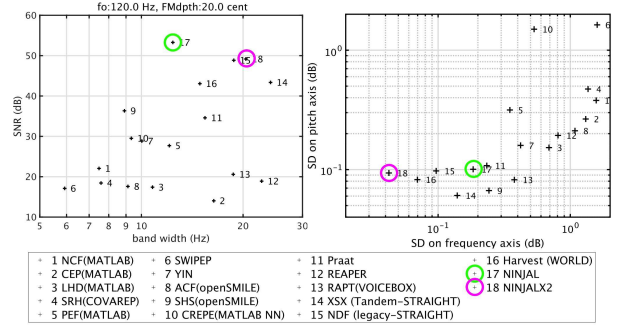}\\
\vspace{-2mm}
\caption{Scatter plot of bandwidth and SNR (LTI power to TD ratio, defined below bw) of each pitch extractor (left plot) and standar deviations (SDs) of gain differences on the modulation frequency and on the fundamental frequency (right plot).
Green and purple circles indicate before and after tuning, NINJAL, and NINJALX2, respectively.}
\label{fig:scatterBwSNR}
\end{center}
\vspace{-6mm}
\end{figure}
Figure~\ref{fig:scatterBwSNR} shows scatter plots of the extractors on the bandwidth-SNR plane for 240~Hz in the left plot.
We noticed that gains of some pitch extractors are not smooth and change depending on both the modulation frequency and the fundamental frequency.
These changes are not desirable for measuring equipment.
The right plot shows the scatter plot of these changes.
The extractors in the bottom left are the better ones.
The movie and these plots indicate that the revised version of NINJAL (NINJALX2) best fits scientific research for voice fundamental frequencies.


\section{Conclusion}
\label{sec:conclusion}
\vspace{-2mm}
We proposed an objective and informative measurement method of pitch extractors' response to frequency-modulated tones.
It uses a measurement procedure based on a new member of time-stretched pulses and simultaneously yields the modulation frequency response and the total distortion.
We applied various pitch extractors and placed them on the bandwidth-SNR plane.
We also applied the proposed method to tune one of the extractors to make it best fit scientific research of voice fundamental frequencies.
We placed the code and the data on our GitHub repository\cite{kawahara2020git} and made it open-sourced.
It consists of example MATLAB functions for applying this method to openSMILE, Praat, and other pitch extractors.
The motivation of this research was particular.
However, the outcome is a general-purpose tool applicable to the more comprehensive range of tuning tasks of pitch extractors, for example, noise tolerance, reverberation tolerance, and tolerance to recording conditions.

\section{Acknowledgements}

We appreciate Heiga Zen of Google Brain and Kikuo Maekawa of NINJAL (National Institute for Japanese Language and Linguistics) for inviting the first author for their projects. Participating in their projects enabled us to develop the reference pitch extractor NINJAL. The naming is an acknowledgment for the Center for Corpus Development, NINJAL. This research was also supported by Grants in Aid for Scientific Research (Kakenhi) by JSPS numbers, JP18K00147, JP18K10708, JP19K21618, JP21H03468, JP21H04900, and JP21H00497.

\vspace{-3mm}
{\small
\bibliographystyle{IEEEtran}
\bibliography{kawaharaPitchIS22}
}

\end{document}